\documentclass[12pt]{article}
\usepackage{latexsym}
\usepackage{amsmath,amsfonts}
\usepackage{times}
\allowdisplaybreaks[4]

\catcode`\@=11

\newcount\hour
\newcount\minute
\newtoks\amorpm \hour=\time\divide\hour by 60\minute
=\time{\multiply\hour by 60 \global\advance\minute by-\hour}
\edef\standardtime{{\ifnum\hour<12 \global\amorpm={am}%
        \else\global\amorpm={pm}\advance\hour by-12 \fi
        \ifnum\hour=0 \hour=12 \fi
        \number\hour:\ifnum\minute<10
        0\fi\number\minute\the\amorpm}}
\edef\militarytime{\number\hour:\ifnum\minute<10 0\fi\number\minute}

\def\draftlabel#1{{\@bsphack\if@filesw {\let\thepage\relax
   \xdef\@gtempa{\write\@auxout{\string
      \newlabel{#1}{{\@currentlabel}{\thepage}}}}}\@gtempa
   \if@nobreak \ifvmode\nobreak\fi\fi\fi\@esphack}
        \gdef\@eqnlabel{#1}}
\def\@eqnlabel{}
\def\@vacuum{}
\def\marginnote#1{}
\def\draftmarginnote#1{\marginpar{\raggedright\scriptsize\tt#1}}
\def\draft{
        \pagestyle{plain}
        \overfullrule=2pt
        \oddsidemargin -.5truein
        \def\@oddhead{\sl \phantom{\today\quad\militarytime} \hfil
        \smash{\Large\sl DRAFT} \hfil \today\quad\militarytime}
        \let\@evenhead\@oddhead
        \let\label=\draftlabel
        \let\marginnote=\draftmarginnote
        \def\ps@empty{\let\@mkboth\@gobbletwo
        \def\@oddfoot{\hfil \smash{\Large\sl DRAFT} \hfil}
        \let\@evenfoot\@oddhead}
        \def\@eqnnum{(\theequation)\rlap{\kern\marginparsep\tt\@eqnlabel}%
        \global\let\@eqnlabel\@vacuum}  }

\newcommand{\rf}[1]{(\ref{#1})}
\renewcommand{\theequation}{\thesection.\arabic{equation}}
\renewcommand{\thefootnote}{\fnsymbol{footnote}}
\newcommand{\newsection}{   
\setcounter{equation}{0}\section}

\def\appendix#1{\addtocounter{section}{1}\setcounter{equation}{0}
\renewcommand{\thesection}{\Alph{section}}
\section*{Appendix \thesection\protect\indent \parbox[t]{11.15cm}{#1}}
\addcontentsline{toc}{section}{Appendix \thesection\ \ \ #1}}

\def\oplussm{{\scriptscriptstyle \oplus}}
\def\ominussm{{\scriptscriptstyle \ominus}}

\def\be{\begin{equation}}
\def\ee{\end{equation}}
\def\beq{\begin{eqnarray}}
\def\eeq{\end{eqnarray}}

\def\parline{\,\partial\kern -0.55em /\,\,}

\def\half{{\frac{1}{2}}}

\def\LL{{\cal L}}
\def\MM{{ M}}
\def\NN{{\cal N}}

\def\Nbf{{\bf N}}

\def\ibf{{\bf i}}
\def\iibf{{\bf ii}}
\def\iiibf{{\bf iii}}

\def\alphab{{\bar{\alpha}}}

\def\rhob{{\bar{\rho}}}
\def\etab{{\bar{\eta}}}
\def\zetab{{\bar{\zeta}}}

\def\phik{|\phi\rangle}
\def\phibr{\langle\phi|}
\def\ck{|c\rangle}
\def\cbk{|\bar{c}\rangle}
\def\cbr{\langle c |}
\def\cbbr{\langle \bar{c}|}

\def\Phik{|\Phi\rangle}
\def\Phibr{\langle\Phi|}

\def\xik{|\xi\rangle}
\def\Xik{|\Xi\rangle}

\def\Ism{{\scriptscriptstyle I}}
\def\IIsm{{\scriptscriptstyle II}}
\def\FPsm{{\scriptscriptstyle FP}}

\def\smponetwo{{\scriptscriptstyle [1,2]}}

\def\Awt{\widetilde{A}}

\def\alpar{\alpha\partial}
\def\albpar{{\bar\alpha\partial}}

\def\Ab{\bar{A}}

\def\Gb{\bar{G}}
\def\Lb{\bar{L}}

\def\cb{\bar{c}}
\def\eb{\bar{e}}
\def\fb{\bar{f}}
\def\gb{\bar{g}}
\def\hb{\bar{h}}
\def\lb{\bar{l}}

\def\rb{\bar{r}}

\def\(A)dS{{\rm (A)dS}}

\def\ext{{\rm ext}}
\def\intrm{{\rm int}}
\def\gh{{\rm gh}}

\def\ssf{{\sf s}}
\def\ssfb{\bar{\sf s}}

\begin{document}


\begin{flushright}
FIAN-TD-2015-12 \hspace{1.8cm} {}~  \\
arXiv: 1511.01836V2 [hep-th] \\
\hspace{3.8cm}  {}~
\end{flushright}

\vspace{1cm}

\begin{center}

{\Large \bf BRST-BV approach to conformal fields }

\vspace{2.5cm}

R.R. Metsaev%
\footnote{ E-mail: metsaev@lpi.ru
}

\vspace{1cm}

{\it Department of Theoretical Physics, Lebedev Physical Institute, \\
  Leninsky prospect 53, Moscow 119991, Russia }

\vspace{3.5cm}

{\bf Abstract}

\end{center}

Using the BRST--BV approach, we consider totally symmetric arbitrary
integer spin conformal fields propagating in flat space.  For such
fields, we obtain the ordinary-derivative BRST--BV Lagrangian that is
invariant under gauge transformations.  In our approach, the
ordinary-derivative Lagrangian and gauge transformations are
constructed in terms of the respective traceless gauge fields and
traceless gauge transformation parameters.  We also obtain a
realization of conformal algebra symmetries on the space of fields and
antifields entering the BRST--BV formulation of conformal fields.

\newpage
\renewcommand{\thefootnote}{\arabic{footnote}}
\setcounter{footnote}{0}

\newsection{  Introduction }

The BRST approach \cite{Becchi:1974xu} is a powerful method of modern
quantum field theory. In gauge field theory, this approach turns out
to be very convenient for studying ultraviolet divergences and the
renormalization procedure. Nowadays, the BRST approach is the main
method for finding the relativistic invariant $S$-matrix. We also note
that the use of the BRST approach for studying string field theory led
to an interesting new development of this approach
\cite{Siegel:1984wx}. Namely, extended versions of the BRST approach
involving antifields were developed and it was demonstrated that such
extended versions turn out to be not only efficient approaches for
studying quantum field theory but also powerful approaches for
constructing theories of classical gauge fields.  The BRST approach
that involves antifields \cite{Batalin:1981jr} will be referred to as
the BRST--BV approach in this paper.

For aesthetic reasons, conformal fields have attracted considerable
interest during a long period of time. The duality conjecture by
Maldacena also triggered interest in studying various aspects of
conformal fields. In the framework of AdS/CFT correspondence,
conformal fields manifest themselves as boundary values of
nonnormalizable solutions of the Dirichlet problem for massless fields
propagating in AdS space.  Plugging a solution of the Dirichlet
problem into the action of a massless AdS field gives a functional of
boundary values of the massless AdS field, which is referred to as the
effective action.  For massless spin-2 and arbitrary integer spin-$s$
fields, it was demonstrated explicitly in the respective
Ref.\cite{Liu:1998bu} and Ref.\cite{Metsaev:2009ym} that the
ultraviolet divergence of the effective action turns out to be a
respective higher-derivative action for conformal spin-2 and for
arbitrary integer spin-$s$ fields. Interesting interrelations between
partition functions of AdS fields and partition functions of conformal
fields are discussed in Ref.\cite{Tseytlin:2013jya}. Taking the
important role played by conformal fields in the AdS/CFT
correspondence into account, we think that study of these fields by
using various approaches is well motivated. In this respect, we note
then that the BRST--BV Lagrangian formulation of conformal fields has
not been discussed previously in the literature.  The major aim of
this paper is to develop a BRST--BV Lagrangian formulation for totally
symmetric arbitrary integer spin conformal fields.%
\footnote{In the literature, totally symmetric conformal fields are
  sometimes referred to as Fradkin--Tseytlin conformal fields.}

The BRST--BV formulation of conformal fields we develop in this paper
is closely related to the ordinary-derivative metric-like formulation
of conformal fields. The ordinary-derivative metric-like formulation
of conformal fields in terms of double-traceless fields was developed
in Refs.\cite{Metsaev:2007fq,Metsaev:2007rw}. In the BRST--BV
framework, we prefer to deal with traceless fields. A formulation of
conformal fields in terms of traceless fields can easily be obtained
from the formulation in terms of double-traceless fields. Therefore,
for the convenience of the reader, we start in Section \ref{sec-01}
with a brief review of our ordinary-derivative metric-like formulation
of conformal fields in terms of the double-traceless fields developed
in Ref.\cite{Metsaev:2007rw} and then we demonstrate how such a
formulation can be used to derive the ordinary-derivative metric-like
formulation of conformal fields in terms of traceless fields.  In
Section \ref{section-03}, we develop a BRST--BV formulation of
conformal fields.  We start with the description of the field content
entering our BRST--BV formulation. After that, we present our result
for the ordinary-derivative BRST--BV Lagrangian of conformal
fields. In our approach to conformal fields, only symmetries of the
Lorentz algebra are realized manifestly. Therefore, in Section
\ref{section-04}, we discuss a realization of conformal symmetries in
the framework of our ordinary-derivative approach.  Namely, we start
with a brief review of a realization of the conformal algebra
symmetries on the space of double-traceless fields obtained in
Ref.\cite{Metsaev:2007rw} and then demonstrate how this realization
can be used to derive the realization of conformal algebra symmetries
on the space of traceless fields.  After that, we present our result
for a realization of conformal algebra symmetries on the space of the
fields and antifields entering our ordinary-derivative BRST--BV
formulation of conformal fields.  In Section \ref{section-05}, we
discuss directions for future research.

\newsection{Ordinary-derivative metric-like formulation of conformal fields}\label{sec-01}

{\bf Field content}. We start with a discussion of the metric-like
formulation of conformal fields in terms of double-traceless fields
developed in Ref.\cite{Metsaev:2007rw}.  To streamline the
presentation of the field content entering the metric-like formulation
of conformal fields, we use oscillators $\alpha^a$, $\alpha^\oplussm$,
$\alpha^\ominussm$, $\zeta$. The oscillators $\alpha^a$ transform as
vectors under the Lorentz algebra $so(d-1,1)$, while the oscillators
$\alpha^\oplussm$, $\alpha^\ominussm$, $\zeta$ transform as scalars
under the Lorentz algebra. Using the oscillators, we note that the
fields entering the metric-like formulation of a conformal field can
be collected into a ket-vector defined by
\be \label{28102015-man03-14}
\phik = \phi(x,\alpha,\alpha^\oplussm,\alpha^\ominussm,\zeta)|0\rangle\,,
\ee
where the argument $\alpha$ in \rf{28102015-man03-14} stands for the
vector oscillators $\alpha^a$, and $x$ stands for coordinates $x^a$ of
the $R^{d-1,1}$ space--time. We also note that the usual fields, which
depend on space-time coordinates $x^a$, can be obtained by expanding
the field $\phi$ \rf{28102015-man03-14} in the oscillators $\alpha^a$,
$\alpha^\oplussm$, $\alpha^\ominussm$, $\zeta$. For the discussion of
a spin-$s$ conformal field, we impose the following algebraic
constraints on the ket-vector $\phik$ \rf{28102015-man03-14}:
\beq
\label{28102015-man03-15} && (N_\alpha + N_\zeta  - s)\phik = 0 \,,
\\
\label{28102015-man03-16} && (N_\zeta + N_{\alpha^\oplussm}+ N_{\alpha^\ominussm} - k_s) \phik = 0 \,,
\\
\label{28102015-man03-17} && (\alphab^2)^2\phik=0\,,
\\
\label{28102015-man03-17-a1} && \hspace{1.5cm} k_s \equiv s + \frac{d-6}{2}\,.
\eeq
The definition of the operators $N_\alpha$, $N_{\alpha^\oplussm}$,
$N_{\alpha^\ominussm}$, $N_\zeta$, $\alphab^2$ appearing in relations
\rf{28102015-man03-15}--\rf{28102015-man03-17} can be found in the
Appendix (see relations
\rf{03112015-man03-39-a2}--\rf{03112015-man03-39-a4}). From constraint
\rf{28102015-man03-15}, we learn that the ket-vector $\phik$ is a
degree-$s$ homogeneous polynomial in the oscillators $\alpha^a$,
$\zeta$, while constraint \rf{28102015-man03-16} tells us that the
ket-vector $\phik$ is a degree-$k_s$ homogeneous polynomial in the
oscillators $\alpha^\oplussm$, $\alpha^\ominussm$, $\zeta$. From
\rf{28102015-man03-17} we learn that the usual tensor fields, which
are obtainable by expanding the field $\phi$ \rf{28102015-man03-14} in
the vector oscillators $\alpha^a$, are nothing but double-traceless
tensor fields of the Lorentz algebra $so(d-1,1)$.

By using constraints \rf{28102015-man03-15}, \rf{28102015-man03-16},
we can illustrate the field content of the ket-vector $\phik$ \rf{28102015-man03-14}
explicitly.  For this, we note that the general
expression for $\phik$ that satisfies algebraic constraints
\rf{28102015-man03-15}, \rf{28102015-man03-16} can be represented as
\beq
\label{28102015-man03-18}  && \hspace{-1cm} \phik \equiv \sum_{s'=0}^s
\frac{\zeta^{s-s'}}{\sqrt{(s-s')!}}|\phi^{s'}\rangle \,,
\\
\label{28102015-man03-19} && |\phi^{s'}\rangle \equiv   \sum_{k'\in
[k_{s'}]_2} \frac{1}{s'!(\frac{k_{s'} + k'}{2})!}\alpha^{a_1} \ldots
\alpha^{a_{s'}} (\alpha^\ominussm)^{^{\frac{k_{s'}+k'}{2}}}
(\alpha^\oplussm)^{^{\frac{k_{s'} - k'}{2}}} \,
\phi_{k'}^{a_1\ldots a_{s'}}|0\rangle\,,
\\
\label{28102015-man03-20} && |\phi^0\rangle \equiv 0 \,, \qquad \hbox{ for } d=4\,,
\\
\label{28102015-man03-21} && \hspace{2cm} k_{s'} \equiv s' + \frac{d-6}{2}\,,
\eeq
where, in \rf{28102015-man03-19} and below, the notation
$ k \in [n]_2$ implies that
\be
k = -n,-n+2,\ldots , n-2,n\,.
\ee
In \rf{28102015-man03-19}, $\phi_{k'}^{a_1\ldots a_{s'}}$ stands for a
rank-$s'$ totally symmetric tensor field of the Lorentz algebra
$so(d-1,1)$. Constraint \rf{28102015-man03-17} implies that the tensor
field $\phi_{k'}^{a_1\ldots a_{s'}}$ is double-traceless,
$\phi_{k'}^{aabb a_5\ldots a_{s'}} =0$, $s'\geq 4$.  We also note that
the subscript $k'$ is used to indicate conformal dimensions of fields,
\be
\label{28102015-man03-25} \Delta(\phi_{k'}^{a_1\ldots a_{s'}}) =
\frac{d-2}{2} + k'\,.
\ee
In the case $d\geq 6$, scalar, vector, and tensor fields appearing in
\rf{28102015-man03-18}, \rf{28102015-man03-19} can be represented as
\beq
\label{28102015-man03-22} &&\phi_{k'}^{a_1\ldots a_s}\,, \hspace{3cm} k'\in [k_s]_2\,;
\\
&&\phi_{k'}^{a_1\ldots a_{s-1}}\,,
\hspace{2.7cm}
k' \in [k_s-1]_2\,;
\\
&&\ldots\ldots
\hspace{3.4cm}
\ldots \ldots\ldots
\nonumber\\[-10pt]
&&\ldots\ldots
\hspace{3.4cm}
\ldots \ldots\ldots
\nonumber\\
\label{28102015-man03-23} && \phi_{k'}^a\,,
\hspace{3.8cm}
k' \in [k_s-s+1]_2\,;
\\
\label{28102015-man03-24} && \phi_{k'}\,,
\hspace{3.8cm}
k' \in [k_s-s]_2\,.
\eeq
We note that scalar fields are collected into the ket-vector
$|\phi^0\rangle$. From \rf{28102015-man03-20}, we see that for $d=4$,
scalar fields do not enter the field content. Therefore, for $d = 4$,
the field content is given in
\rf{28102015-man03-22}-\rf{28102015-man03-23}. In other words, scalar
fields $\phi_{k'}$ enter the field content only when $d\geq 6$.

To illustrate the field content presented in
\rf{28102015-man03-22}-\rf{28102015-man03-24} more explicitly, we use
the shorthand notation $\phi_{k'}^{s'}$ for the field
$\phi_{k'}^{a_1\ldots a_{s'}}$.  We then note that the fields in
\rf{28102015-man03-22}-\rf{28102015-man03-24} can be represented as
{\small
\beq
&& \hspace{2.5cm} \hbox{Field content for} \ \ d \geq 6\,,\quad s -
\hbox{ arbitrary integer}
\nonumber\\
&&\hspace{-1cm}
\phi_{-k_s}^s
\hspace{1cm} \phi_{2-k_s}^s
\hspace{1cm} \ldots
\hspace{1cm}\ldots
\hspace{1cm} \ldots
\hspace{1cm} \ldots
\hspace{1cm}\ldots
\hspace{1cm} \phi_{k_s-2}^s
\hspace{1cm}  \phi_{k_s}^s \qquad
\nonumber\\[15pt]
&& \hspace{-0.3cm} \phi_{1-k_s}^{s-1}
\hspace{1cm}\phi_{3-k_s}^{s-1}
\hspace{1cm} \ldots
\hspace{1cm} \ldots
\hspace{1cm}\ldots
\hspace{1cm}\ldots
\hspace{1cm}\phi_{k_s-3}^{s-1}
\hspace{1cm} \phi_{k_s-1}^{s-1}
\nonumber\\[14pt]
&& \hspace{0.5cm}
\ldots \hspace{1cm}  \ldots
\hspace{1cm}  \ldots \hspace{1cm} \ldots
\hspace{1cm} \ldots   \hspace{1cm} \ldots
\hspace{1cm} \ldots \hspace{1cm} \ldots
\\[14pt]
&& \hspace{0.8cm}  \phi_{s-1-k_s}^1
\hspace{1cm} \phi_{s+1-k_s}^1
\hspace{1cm}  \ldots  \hspace{1cm} \ldots
\hspace{1cm} \phi_{k_s-s-1}^1
\hspace{0.8cm} \phi_{k_s-s+1}^1
\nonumber\\[15pt]
&& \hspace{1.8cm} \phi_{s-k_s}^0
\hspace{1.2cm} \phi_{s-k_s+2}^0
\hspace{1.2cm} \ldots
\hspace{1cm} \phi_{k_s-s-2}^0
\hspace{1cm} \phi_{k_s-s}^0
\nonumber
\eeq }

We now discuss representations for the Lagrangian in terms of
double-traceless and traceless fields in turn.

\medskip
\noindent{\bf Representation for the Lagrangian in terms of
  double-trace fields}. An ordinary-derivative representation for the
action and the Lagrangian of a spin-$s$ conformal field in terms of
double-traceless fields described above was obtained in
Ref.\cite{Metsaev:2007rw}:
\beq
&& S = \int d^d x\, \LL\,,
\\
\label{28102015-man03-01}
\LL  & = &
\frac{1}{2}
\phibr (1-\frac{1}{4}\alpha^2\alphab^2)
(\Box - M^{\oplussm\oplussm})
\phik
+ \half
\langle \Lb
\phi|
\Lb\phi\rangle \,,\qquad
\\
\label{28102015-man03-02} && M^{\oplussm\oplussm} \equiv \alpha^\oplussm  \bar\alpha^\oplussm\,,\qquad
\\
\label{28102015-man03-03} &&   \Lb \equiv \albpar - \half \alpar \bar\alpha^2 -
\Pi^\smponetwo \eb_1 + \half e_1 \bar\alpha^2\,,\qquad
\\
\label{28102015-man03-04} &&  e_1 \equiv \zeta e_\zeta \bar\alpha^\oplussm\,,\qquad \qquad \eb_1 \equiv - \alpha^\oplussm
e_\zeta \bar\zeta\,, \qquad e_\zeta \equiv \Bigl(\frac{2s+d-4-N_\zeta}{2s+d-4-2N_\zeta}\Bigr)^{1/2}\,, \qquad
\\
\label{28102015-man03-05} && \Pi^\smponetwo \equiv 1 - \alpha^2\frac{1}{2(2N_\alpha+d)}\bar\alpha^2\,.
\eeq
In \rf{28102015-man03-01}, the bra-vector is defined in accordance
with the rule $\phibr \equiv (\phik)^\dagger$. We also use the
conventions that $|\Lb\phi\rangle \equiv \Lb \phik$ and
$\langle \Lb\phi| \equiv (|\Lb\phi\rangle)^\dagger$. The remaining
notation can be found in the Appendix (see relations
\rf{03112015-man03-39-a1}-\rf{03112015-man03-39-a4}).

We refer to the quantity $|\Lb\phi\rangle$ as the conformal de Donder
divergence.  We recall that the standard de Donder divergence, which
enters the massless field Lagrangian in $R^{d-1,1}$, is obtained by
setting $e_1=0$, $\eb_1=0$ in \rf{28102015-man03-03} (see, e.g.,
Ref.\cite{Guttenberg:2008qe}). From \rf{28102015-man03-01}, we see
that it is the use of the conformal de Donder divergence that allows
us to considerably simplify the representation for the gauge invariant
Lagrangian of a conformal field.%
\footnote{ The use of modified de Donder divergences for studying the
  metric-like formulation of fields in AdS space can be found in
  Refs.\cite{Metsaev:2008ks,Metsaev:2009hp}. For a massive field in
  $R^{d-1,1}$, the modified de Donder divergence was found in
  Ref.\cite{Metsaev:2008fs}.  }

To describe gauge symmetries of Lagrangian \rf{28102015-man03-01}, we
note that appropriate gauge transformation parameters can be collected
into the ket-vector
\be \label{29102015-man03-02}
\xik = \xi(x,\alpha,\alpha^\oplussm,\alpha^\ominussm,\zeta)|0\rangle\,,
\ee
which satisfies the algebraic constraints
\beq
\label{29102015-man03-03} && (N_\alpha + N_\zeta - s +1 ) \xik=0\,,
\\
\label{29102015-man03-04} && (N_\zeta +  N_{\alpha^\oplussm}+ N_{\alpha^\ominussm} - k_s ) \xik=0 \,,
\\
\label{29102015-man03-05} && \bar\alpha^2 \xik=0 \,,
\eeq
where $k_s$ is defined in \rf{28102015-man03-17-a1}. From
\rf{29102015-man03-03}, we learn that the ket-vector $\xik$ is a
degree-$(s-1)$ homogeneous polynomial in $\alpha^a$, $\zeta$, while
from \rf{29102015-man03-04}, we learn that the ket-vector $\xik$ is a
degree-$k_s$ homogeneous polynomial in $\zeta$, $\alpha^\oplussm$,
$\alpha^\ominussm$. Taking this into account, we see that the
ket-vector $\xik$ can be expanded in the oscillators as follows
\beq
\label{29102015-man03-06} \xik & = & \sum_{s'=0}^{s-1}
\frac{\zeta^{s-1-s'}}{\sqrt{(s-1-s')!}}|\xi^{s'}\rangle \,,
\\
\label{29102015-man03-07} && |\xi^{s'}\rangle \equiv  \sum_{k'\in
[k_{s'}+1]_2}\frac{1}{s'!(\frac{k_{s'}+1+k'}{2})!}\alpha^{a_1}
\ldots \alpha^{a_{s'}}
(\alpha^\ominussm)^{^{\frac{k_{s'}+1+k'}{2}}}
(\alpha^\oplussm)^{^{\frac{k_{s'}+1-k'}{2}}} \,
\xi_{k'-1}^{a_1\ldots a_{s'}} |0\rangle\,,\qquad
\eeq
where $k_{s'}$ is defined in \rf{28102015-man03-21}. In
\rf{29102015-man03-07}, the gauge transformation parameter
$\xi_{k'}^{a_1\ldots a_{s'}}$ is a rank-$s'$ totally symmetric tensor
field of the Lorentz algebra $so(d-1,1)$. Constraint
\rf{29102015-man03-05} tells us that the gauge transformation
parameter $\xi_{k'}^{a_1\ldots a_{s'}}$ is traceless,
$\xi_{k'}^{aa a_3\ldots a_{s'}} =0$, $s'\geq 2$.  The conformal
dimension of the gauge transformation parameter
$\xi_{k'}^{a_1\ldots a_{s'}}$ is given by
\be \label{29102015-man03-08}
\Delta(\xi_{k'}^{a_1\ldots a_{s'}})
=
\frac{d-2}{2}
+ k'\,.
\ee

Using ket-vectors $\phik$ \rf{28102015-man03-18} and $\xik$
\rf{29102015-man03-06}, gauge transformations can be written as
\be \label{29102015-man03-09}
\delta \phik
=
G \xik  \,,
\qquad
G \equiv
\alpar
- e_1
- \alpha^2 \frac{1}{2N_\alpha +d-2}
\eb_1\,,
\ee
where the operators $e_1$, $\eb_1$ are defined in
\rf{28102015-man03-04}.

\noindent{\bf Representation for the Lagrangian in terms of traceless
  fields}. A Lagrangian formulation of a conformal field in terms of
traceless fields can easily be obtained by using the formulation in
terms of double-traceless fields discussed above.  For this, we note
that the double-traceless ket-vector $\phik$ \rf{28102015-man03-14},
\rf{28102015-man03-17} can be decomposed into two traceless
ket-vectors, denoted by $|\phi_\Ism \rangle$, $|\phi_\IIsm \rangle$, as
\be
\label{28102015-man03-06} \phik = |\phi_\Ism \rangle   + \alpha^2 \NN |\phi_\IIsm \rangle\,,
\hspace{1.5cm} \NN\equiv \bigl((2N_\alpha+d)(2N_\alpha+d-2)\bigr)^{-1/2}\,,
\ee
where the tracelessness constraint for the ket-vectors
$|\phi_\Ism \rangle$, $|\phi_\IIsm \rangle$ takes the form
\be \label{28102015-man03-08}
\bar\alpha^2 |\phi_\Ism \rangle =0\,,  \qquad \bar\alpha^2 |\phi_\IIsm \rangle =0\,.
\ee
Plugging \rf{28102015-man03-06} in \rf{28102015-man03-01}, we obtain a
representation for the Lagrangian in terms of traceless ket-vectors,
\be \label{28102015-man03-11}
\LL  =    \half \langle \phi_\Ism | (\Box - M^{\oplussm\oplussm})  |\phi_\Ism \rangle - \half \langle \phi_\IIsm | (\Box - M^{\oplussm\oplussm})  |\phi_\IIsm \rangle + \half \langle \Lb \phi|\Lb\phi\rangle \,,
\ee
where the operator $M^{\oplussm\oplussm}$ takes the same form as in
\rf{28102015-man03-02}. The de Donder divergence $|\Lb\phi\rangle$
appearing in \rf{28102015-man03-11} is represented in terms of the
traceless ket-vectors $|\phi_\Ism\rangle$, $|\phi_\IIsm\rangle$ as
\beq
\label{28102015-man03-11-b0} \Lb\phik & = &  \Lb_\Ism |\phi_\Ism\rangle + L_\IIsm |\phi_\IIsm\rangle\,,
\\
\label{28102015-man03-11-b1} && \Lb_\Ism \equiv \albpar - \eb_1\,,
\\
 \label{28102015-man03-11-b2} && L_\IIsm \equiv - A^a\partial^a f_{1\zeta}  + e_1 f_{1\zeta}^{-1}\,,
\\
 \label{28102015-man03-11-b3} && f_{1\zeta} \equiv \Bigl(\frac{2s+d-6-2N_\zeta}{2s+d-4-2N_\zeta}\Bigr)^{1/2}\,,
\\
 \label{28102015-man03-11-b4} && A^a \equiv \alpha^a - \alpha^2 \frac{1}{2N_\alpha+d}\bar\alpha^a\,,
\eeq
where $e_1$, $\eb_1$ are given in \rf{28102015-man03-04}. In terms of
the traceless ket-vectors $|\phi_\Ism\rangle$, $\phi_\IIsm\rangle$,
gauge transformations \rf{29102015-man03-09} can be represented as
\beq
\label{30102015-man03-01} && \delta |\phi_\Ism\rangle = G_\Ism \xik\,,
\\
\label{30102015-man03-02} && \delta |\phi_\IIsm\rangle = \Gb_\IIsm \xik\,,
\\
\label{30102015-man03-02-b1} && \hspace{1.5cm} G_\Ism \equiv A^a\partial^a - e_1\,,
\\
\label{30102015-man03-02-b2} && \hspace{1.5cm} \Gb_\IIsm \equiv f_{1\zeta} \albpar - f_{1\zeta}^{-1}\eb_1\,,
\eeq
where the gauge transformation parameter $\xik$ takes the same form as
in \rf{29102015-man03-06}.  Gauge transformations
\rf{30102015-man03-01}, \rf{30102015-man03-02} can easily be obtained
from the ones in \rf{29102015-man03-09} by noticing that the inverse
to relation in \rf{28102015-man03-06} takes the form
\beq
\label{28102015-man03-09} && |\phi_\Ism\rangle = \Pi^\smponetwo \phik\,,
\\
\label{28102015-man03-10} && |\phi_\IIsm\rangle = \half \Bigl(\frac{2N_\alpha +d-2}{2N_\alpha+d}\Bigr)^{1/2}
\alphab^2 \phik\,,
\eeq
where the operator $ \Pi^\smponetwo$ is defined in
\rf{28102015-man03-05}.

From the presentation given above, we see that our formulations in
terms of double-traceless and traceless fields are completely
equivalent.

\newsection{BRST--BV Lagrangian of conformal fields}\label{section-03}

{\bf Field content}. In order to streamline the presentation of the
field content entering our BRST--BV formulation of a conformal field,
we introduce a Grassmann coordinate $\theta$, $\theta^2=0$,
Grassmann-even oscillators $\alpha^a$, $\alpha^\oplussm$,
$\alpha^\ominussm$, $\zeta$, and Grassmann-odd oscillators $\eta$,
$\rho$. The Grassmann-even oscillators $\alpha^a$ transform as vectors
under the Lorentz algebra $so(d-1,1)$, while the Grassmann coordinate
$\theta$, the Grassmann-even oscillators $\alpha^\oplussm$,
$\alpha^\ominussm$, $\zeta$, and the Grassmann-odd oscillators $\eta$,
$\rho$ transform as scalars under the Lorentz algebra. Using the
Grassmann coordinate $\theta$ and the oscillators, the fields entering
our BRST--BV formulation of a conformal field can be collected into a
ket-vector defined by
\be \label{27-10-2015-manus03-01}
\Phik = \Phi(x,\theta,\alpha,\alpha^\oplussm,\alpha^\ominussm,\zeta,\eta,\rho)|0\rangle\,,
\ee
where the argument $\alpha$ in \rf{27-10-2015-manus03-01} stands for
the vector oscillators $\alpha^a$, while the argument $x$ stands for
coordinates $x^a$ of the $R^{d-1,1}$ space--time. By definition, the
field $\Phi$ in \rf{27-10-2015-manus03-01} is considered to be
Grassmann even. We also note that usual fields, which depend on the
space--time coordinates $x^a$, can be obtained by expanding the field
$\Phi$ \rf{27-10-2015-manus03-01} in the $\theta$ and the
above-introduced oscillators $\alpha^a$, $\alpha^\oplussm$,
$\alpha^\ominussm$, $\zeta$, $\eta$, $\rho$.

For the discussion of a spin-$s$ conformal field, we impose the
following algebraic constraints on the ket-vector $\Phik$
\rf{27-10-2015-manus03-01}:
\beq
\label{27-10-2015-manus03-02} && (N_\alpha + N_\zeta + N_\eta + N_\rho - s)\Phik = 0 \,,
\\
\label{27-10-2015-manus03-03} && (N_\zeta + N_{\alpha^\oplussm}+ N_{\alpha^\ominussm} - k_s) \Phik = 0 \,,
\\
\label{27-10-2015-manus03-04} && \alphab^2\Phik=0\,,
\eeq
where $k_s$ is defined in \rf{28102015-man03-17-a1}, while the operators $N_\alpha$, $N_{\alpha^\oplussm}$,
$N_{\alpha^\ominussm}$, $N_\zeta$, $N_\eta$, $N_\rho$, $\alphab^2$
appearing in \rf{27-10-2015-manus03-02}--\rf{27-10-2015-manus03-04}
are defined in the Appendix (see relations
\rf{03112015-man03-39-a2}--\rf{03112015-man03-39-a4}). From constraint
\rf{27-10-2015-manus03-02}, we learn that the ket-vector $\Phik$ is a
degree-$s$ homogeneous polynomial in the oscillators $\alpha^a$,
$\zeta$, $\eta$, $\rho$, while constraint \rf{27-10-2015-manus03-03}
tells us that the ket-vector $\Phik$ is a degree-$k_s$ homogeneous
polynomial in the oscillators $\alpha^\oplussm$, $\alpha^\ominussm$,
$\zeta$.  From constraint \rf{27-10-2015-manus03-04} we learn that the
usual tensor fields, which are obtainable by expanding the field
$\Phi$ \rf{27-10-2015-manus03-01} in the vector oscillators
$\alpha^a$, are nothing but traceless tensor fields of the Lorentz
algebra $so(d-1,1)$.

The use of constraints \rf{27-10-2015-manus03-02},
\rf{27-10-2015-manus03-03} allows us to illustrate a catalogue of
scalar, vector, and tensor fields entering the ket-vector $\Phik$
\rf{27-10-2015-manus03-01} in a rather straightforward way. For this,
we note that the expansion of $\Phik$ \rf{27-10-2015-manus03-01} in
the Grassmann coordinate $\theta$ and the Grassmann-odd oscillators
$\eta$, $\rho$ can be presented as
\beq
\label{27-10-2015-manus03-05}
\Phik & = &
\phik
+ \theta
|\phi_*\rangle\,,
\\
\label{27-10-2015-manus03-06} &&
\phik
=
|\phi_\Ism\rangle
+ \rho |c\rangle
+ \eta |\cb\rangle
+ \rho \eta
|\phi_\IIsm\rangle\,,
\\
\label{27-10-2015-manus03-07} &&
|\phi_*\rangle
= |\phi_{*\Ism}\rangle
+ \rho |\cb_*\rangle
+ \eta |c_*\rangle
+ \rho \eta
|\phi_{*\IIsm}\rangle\,.
\eeq
We now note that the ket-vectors appearing in the right-hand side of
\rf{27-10-2015-manus03-06}, \rf{27-10-2015-manus03-07} depend only on
the vector oscillators $\alpha^a$ and the scalar oscillators $\zeta$,
$\alpha^\oplussm$, $\alpha^\ominussm$. In this paper, the ket-vectors
$\phik$ \rf{27-10-2015-manus03-06} and $|\phi_*\rangle$
\rf{27-10-2015-manus03-07} are referred to as the respective
ket-vectors of fields and antifields.%
\footnote{ Our terminology for ket-vectors in
  \rf{27-10-2015-manus03-06}, \rf{27-10-2015-manus03-07} does not
  coincide with the standard terminology used in the
  literature. Matching our simplified terminology and the standard one
  is discussed at the end of the Appendix.  }
We note that the algebraic constraints for the ket-vector of fields
$\phik$ and the ket-vector of antifields $|\phi_*\rangle$ take the
same form as the ones for the $\Phik$ in
\rf{27-10-2015-manus03-02}-\rf{27-10-2015-manus03-04}.  This implies
that expansions of the ket-vector of fields and the ket-vector of
antifields in terms of the respective scalar, vector, and tensor
fields and antifields of the Lorentz algebra $so(d-1,1)$ take the same
form. Therefore, for illustration purpose, we consider an expansion of
the ket-vector of fields $\phik$.  Taking constraints
\rf{27-10-2015-manus03-02}, \rf{27-10-2015-manus03-03} into account,
we verify that the ket-vector $|\phi_{\Ism,\IIsm}\rangle$, $\ck$,
$\cbk$ \rf{27-10-2015-manus03-06} depends on the oscillators
$\alpha^a$, $\alpha^\oplussm$, $\alpha^\ominussm$, $\zeta$ as
\beq
\label{27-10-2015-manus03-08} && \hspace{-1.3cm} |\phi_\Ism\rangle = \sum_{s'=0}^s \frac{\zeta^{s-s'}}{\sqrt{(s-s')!}} |\phi_\Ism^{s'}\rangle \,,
\nonumber\\
&& |\phi_\Ism^{s'}\rangle \equiv   \sum_{k'\in
[k_{s'}]_2} \frac{1}{s'!(\frac{k_{s'} + k'}{2})!}\alpha^{a_1} \ldots
\alpha^{a_{s'}} (\alpha^\ominussm)^{^{\frac{k_{s'}+k'}{2}}}
(\alpha^\oplussm)^{^{\frac{k_{s'} - k'}{2}}} \,
\phi_{\Ism,k'}^{a_1\ldots a_{s'}}|0\rangle\,,
\\
\label{27-10-2015-manus03-09} && \hspace{-1.3cm} \ck = \sum_{s'=0}^{s-1} \frac{\zeta^{s-1-s'}}{\sqrt{(s-1-s')!}} |c^{s'}\rangle \,, %
\nonumber\\
&& |c^{s'}\rangle \equiv   \sum_{k'\in
[k_{s'}+1]_2} \frac{1}{s'!(\frac{k_{s'}+1 + k'}{2})!}\alpha^{a_1} \ldots
\alpha^{a_{s'}} (\alpha^\ominussm)^{^{\frac{k_{s'} + 1 + k'}{2}}}
(\alpha^\oplussm)^{^{\frac{k_{s'} + 1 - k'}{2}}} \,
c_{k'}^{a_1\ldots a_{s'}}|0\rangle\,,
\\
\label{27-10-2015-manus03-10}&& \hspace{-1.3cm}\cbk = \sum_{s'=0}^{s-1} \frac{\zeta^{s-1-s'}}{\sqrt{(s-1-s')!}} |\cb^{s'}\rangle \,,
\nonumber\\
&& |\cb^{s'}\rangle \equiv   \sum_{k'\in
[k_{s'}+1]_2} \frac{1}{s'!(\frac{k_{s'}+1 + k'}{2})!}\alpha^{a_1} \ldots
\alpha^{a_{s'}} (\alpha^\ominussm)^{^{\frac{k_{s'} + 1 + k'}{2}}}
(\alpha^\oplussm)^{^{\frac{k_{s'} + 1 - k'}{2}}} \,
\cb_{k'}^{a_1\ldots a_{s'}}|0\rangle\,,
\\
\label{27-10-2015-manus03-11} && \hspace{-1.3cm} |\phi_\IIsm\rangle = \sum_{s'=0}^{s-2} \frac{\zeta^{s-2-s'}}{\sqrt{(s-2-s')!}} |\phi_\IIsm^{s'}\rangle \,,
\nonumber\\
&& |\phi_{\IIsm}^{s'}\rangle \equiv   \sum_{k'\in
[k_{s'}+2]_2} \frac{1}{s'!(\frac{k_{s'}+2 + k'}{2})!}\alpha^{a_1} \ldots
\alpha^{a_{s'}} (\alpha^\ominussm)^{^{\frac{k_{s'} + 2 + k'}{2}}}
(\alpha^\oplussm)^{^{\frac{k_{s'} + 2 - k'}{2}}} \,
\phi_{\IIsm,k'}^{a_1\ldots a_{s'}}|0\rangle\,, \qquad
\eeq
where, in \rf{27-10-2015-manus03-08}, $|\phi_\Ism^0\rangle = 0$ if
$d=4$. The $k_{s'}$ appearing in
\rf{27-10-2015-manus03-08}--\rf{27-10-2015-manus03-11} is defined in
\rf{28102015-man03-21}.

Relations \rf{27-10-2015-manus03-08}--\rf{27-10-2015-manus03-11}
explicitly illustrate the field content entering the ket-vector of
fields \rf{27-10-2015-manus03-06}. We see a set of fields denoted by
$\phi_{\Ism,k'}^{a_1\ldots a_{s'}}$, $c_{k'}^{a_1\ldots a_{s'}}$,
$\cb_{k'}^{a_1\ldots a_{s'}}$, $\phi_{\IIsm,k'}^{a_1\ldots a_{s'}}$.
Fields with the values $s'=0$, $s'=1$, and $s'\geq 2$ correspond to
the respective scalar, vector, and tensor fields of the Lorentz
algebra $so(d-1,1)$. All the tensor fields are totally symmetric
tensor fields.  From constraint \rf{27-10-2015-manus03-04}, we learn
that all the tensor fields are realized as traceless tensors of the
Lorentz algebra $so(d-1,1)$.  As is well known, in the BRST--BV
approach, each field enters the game with a corresponding
antifield. This implies that in our approach, a catalogue of the
antifields is described by the antifields denoted by
$\phi_{*\Ism,k'}^{a_1\ldots a_{s'}}$, $c_{*k'}^{a_1\ldots a_{s'}}$,
$\cb_{*k'}^{a_1\ldots a_{s'}}$, $\phi_{*\IIsm,k'}^{a_1\ldots a_{s'}}$.
We also note that ket-vectors of antifields $|\phi_{*\Ism}\rangle$,
$|c_*\rangle$, $|\cb_*\rangle$, $|\phi_{*\IIsm}\rangle$ given in
\rf{27-10-2015-manus03-07} are expressed in terms of the antifields
$\phi_{*\Ism,k'}^{a_1\ldots a_{s'}}$, $c_{*k'}^{a_1\ldots a_{s'}}$,
$\cb_{*k'}^{a_1\ldots a_{s'}}$, $\phi_{*\IIsm,k'}^{a_1\ldots a_{s'}}$
in the same way as in
\rf{27-10-2015-manus03-08}--\rf{27-10-2015-manus03-11}.  Thus, the
ket-vector $\Phik$ describes fields and antifields that are realized
as scalars, vectors, and totally symmetric traceless tensors of the
Lorentz algebra $so(d-1,1)$.

\bigskip
\noindent {\bf BRST--BV Lagrangian of conformal fields}. A general
representation for the BRST--BV action that describes the dynamics of
fields and antifields in the $R^{d-1,1}$ space-time takes the form
\cite{Siegel:1984wx}
\beq
\label{30102015-man03-03}
&& S
= \int d^dx\,
\LL\,,
\\
\label{30102015-man03-04} && \LL =
\half
\int d\theta
\Phibr
Q \Phik\,.
\eeq
The BRST operator $Q$ appearing in expression \rf{30102015-man03-04}
for the Lagrangian admits the general representation
\be \label{30102015-man03-05}
Q
=
\theta (\Box - \MM^{\oplussm\oplussm} )
+ M^{\eta a} \partial^a
+ M^{\oplussm\eta}
+ M^{\eta\eta} \partial_\theta\,,
\ee
where the notation $\partial_\theta$ is used for the left derivative
of the Grassmann coordinate,
$\partial_\theta = \partial/\partial\theta$, while the notation
$\Box=\partial^a\partial^a$ is used for the D'Alembert operator in the
space--time $R^{d-1,1}$.

From the expression for the BRST operator given in
\rf{30102015-man03-05}, we see that the BRST operator is defined by
the operators $\MM^{\oplussm\oplussm}$, $M^{\eta a}$,
$M^{\oplussm\eta}$, $M^{\eta\eta}$. In what follows, the operator
$M^{\oplussm\oplussm}$ is referred to as the mass operator, while the
operators $M^{\eta a}$, $M^{\oplussm\eta}$, $M^{\eta\eta}$ are
referred to as spin operators.  The mass and spin operators depend
only on the oscillators, i.e., the mass and spin operators do not
depend on the space-time coordinates $x^a$, on the Grassmann
coordinate $\theta$, and on the derivatives $\partial^a$,
$\partial_\theta$.

The basic equation for the BRST operator $Q^2=0$ amounts to the
following (anti)commutation relations for the mass and spin operators:
\beq
\label{31102015-man03-01} && \{ M^{\eta a},M^{\eta b}\} = - 2\eta^{ab} M^{\eta\eta}\,,
\\
\label{31102015-man03-02} && \{ M^{\oplussm \eta} ,M^{\oplussm \eta} \} = 2 M^{\oplussm\oplussm} M^{\eta\eta}\,,
\\
\label{31102015-man03-03} && [M^{\oplussm\oplussm},M^{\eta a}]= 0\,, \hspace{1cm} [M^{\oplussm\oplussm},M^{\oplussm \eta}]= 0\,, \hspace{1.2cm} [M^{\oplussm\oplussm},M^{\eta\eta}]= 0\,,
\\
\label{31102015-man03-04} && \{ M^{\eta a},M^{\oplussm \eta} \}= 0\,, \qquad [M^{\eta a},M^{\eta\eta} ] = 0\,, \hspace{1.3cm} [M^{\oplussm \eta},M^{\eta\eta}]= 0\,.
\eeq

BRST--BV action \rf{30102015-man03-03} is invariant under the gauge
transformations given by
\be  \label{31102015-man03-05}
\delta \Phik = Q \Xik,
\ee
where $\Xik$ stands for a ket-vector of the gauge transformation
parameters. The ket-vector $\Xik$ can be represented as
\be \label{31102015-man03-06}
\Xik =  \Xi(x,\theta,\alpha,\alpha^\oplussm,\alpha^\ominussm,\zeta,\eta,\rho)|0\rangle\,,
\ee
where the arguments $x$ and $\alpha$ in \rf{31102015-man03-06} stand
for the respective coordinates $x^a$ of the space--time $R^{d-1,1}$
and vector oscillators $\alpha^a$. Usual gauge transformation
parameters, which depend on the space-time coordinates $x^a$, are
obtained by expanding the generating function $\Xi$
\rf{31102015-man03-06} in the $\theta$ and the oscillators $\alpha^a$,
$\alpha^\oplussm$, $\alpha^\ominussm$, $\zeta$, $\eta$, $\rho$. The
generating function $\Xi$ is considered to be Grassmann odd. The
ket-vector $\Xik$ satisfies, by definition, the algebraic constraints
\beq
\label{31102015-man03-07} && (N_\alpha + N_\zeta + N_\eta + N_\rho - s)\Xik = 0 \,,
\\
\label{31102015-man03-08} && (N_\zeta + N_{\alpha^\oplussm}+ N_{\alpha^\ominussm} - k_s) \Xik = 0 \,,
\\
\label{31102015-man03-09} && \alphab^2\Xik=0\,,
\eeq
where $k_s$ is defined in \rf{28102015-man03-17-a1}.

Comparing the algebraic constraints given in
\rf{27-10-2015-manus03-02}--\rf{27-10-2015-manus03-04} and
\rf{31102015-man03-07}--\rf{31102015-man03-09}, we see that
ket-vectors $\Phik$ and $\Xik$ satisfy the same constraints.
Therefore, using the results of our analysis of the constraints for
the ket-vector of gauge fields $\Phik$, we conclude the following:

\noindent \ibf) The ket-vector of gauge transformation parameters
$\Xik$ is built of gauge transformation parameters that are realized
as scalar, vector, and totally symmetric traceless tensor fields of
the Lorentz algebra $so(d-1,1)$.

\noindent \iibf) An expansion of the ket-vector of the gauge
transformation parameter $\Xik$ in terms of scalar, vector, and tensor
gauge transformation parameters is obtained simply by replacing the
gauge fields in relations
\rf{27-10-2015-manus03-05}--\rf{27-10-2015-manus03-11} by the scalar,
vector and tensor gauge transformation parameters.

It is clear from the discussion given above that all that is required
to find the BRST--BV action and the corresponding gauge
transformations is the BRST operator.  We also see that in order to
build the BRST operator, we should find a realization for the mass and
spin operators on the space of ket-vector $\Phik$ and plug that
realization for the mass and spin operators into expression for the
BRST operator in \rf{30102015-man03-05}. In our approach, a spin-$s$
conformal field is described by the ket-vector $\Phik$ that by
definition, satisfies constraints
\rf{27-10-2015-manus03-02}--\rf{27-10-2015-manus03-04}. This implies
that on space of the ket-vector $\Phik$ subject to algebraic
constraints \rf{27-10-2015-manus03-02}--\rf{27-10-2015-manus03-04}, we
should find a realization of (anti)commutation relations for the mass
and spin operators \rf{31102015-man03-01}-\rf{31102015-man03-04}.  Our
result for the mass and spin operators is as follows%
\footnote{ We note that restrictions imposed by (anti)commutation
relations \rf{31102015-man03-01}--\rf{31102015-man03-04} do not
allow determining the mass and spin operators uniquely.  To fix the
mass and spin operators uniquely, one has to also analyse the
restrictions imposed by conformal algebra symmetries. Such an
analysis can be done by using the method we described in the
framework of the metric-like approach to conformal fields in
Appendix B in Ref.\cite{Metsaev:2007rw}.  A discussion of conformal
symmetries in our BRST--BV approach can be found in
Sec. \ref{section-04} in this paper.  }
\beq
&& M^{\oplussm\oplussm} = \alpha^\oplussm \alphab^\oplussm \,,
\\
&&  M^{\eta a}
= \eta
g_{\rho \zeta}
\alphab^a
+ A^a
\gb_{\eta \zeta}
\etab\,,
\\
&& M^{\eta\eta}
=\eta\etab\,,
\\
&& M^{\oplussm\eta}
= \eta
l_{\rho \zeta}^\oplussm
\zetab
+ \zeta
\lb_{\eta \zeta}^\oplussm
\etab \,,
\eeq
where we use the following notation
\beq
\label{16082015-man-14} && A^a \equiv \alpha^a - \alpha^2\frac{1}{2N_\alpha  + d} \alphab^a\,,
\\
\label{16082015-man-15} && g_{\rho \zeta} \equiv \Bigl(\frac{2s +d-4-2N_\zeta-2N_\rho}{2s +d-4-2N_\zeta}\Bigr)^{1/2} \,,
\\
\label{16082015-man-16} && \gb_{\eta \zeta} \equiv - \Bigl(\frac{2s +d-4-2N_\zeta-2N_\eta}{2s +d-4-2N_\zeta}\Bigr)^{1/2}\,,
\\
\label{16082015-man-17} && l_{\rho \zeta}^\oplussm \equiv \alpha^\oplussm e_\zeta\Bigl(\frac{2s +d-4-2N_\zeta-2N_\rho}{2s +d-4-2N_\zeta}\Bigr)^{-1/2} \,,
\\
\label{16082015-man-18} && \lb_{\eta \zeta}^\oplussm \equiv \alphab^\oplussm e_\zeta\Bigl(\frac{2s +d-4-2N_\zeta-2N_\eta}{2s +d-4-2N_\zeta}\Bigr)^{-1/2}\,,
\\
\label{16082015-man-19} && \hspace{1cm} e_\zeta \equiv \Bigl(\frac{2s +d-4-N_\zeta}{2s +d-4-2N_\zeta}\Bigr)^{1/2}\,,
\eeq
while the definition of the operators $N_\alpha$,
$N_{\alpha^\oplussm}$, $N_{\alpha^\ominussm}$, $N_\zeta$ can be found
in the Appendix (see relations \rf{03112015-man03-39-a3},
\rf{03112015-man03-39-a4}).

\medskip
The following remarks are in order.

\noindent

\noindent
\ibf) From \rf{30102015-man03-05}, we see that our BRST operator is a
degree-2 polynomial in the space--time derivatives $\partial^a$. In
other words, there are no higher-derivative terms in our BRST--BV
Lagrangian formulation of conformal fields. We note then that it is
the field content described in
\rf{27-10-2015-manus03-08}-\rf{27-10-2015-manus03-11} that allows us
to develop the ordinary-derivative BRST--BV Lagrangian formulation of
conformal fields.

\noindent
\iibf) We demonstrate how our BRST--BV Lagrangian is related to the
Lagrangian entering the metric-like formulation in terms of traceless
fields \rf{28102015-man03-11}. For this, we equate all fields and
antifields with a nonzero ghost number to zero:
\be \label{16112015-man-01}
\ck = 0 \,, \qquad \cbk =  0\,, \qquad |\phi_{*\Ism}\rangle = 0 \,, \qquad  |c_*\rangle = 0 \,, \qquad |\phi_{*\IIsm}\rangle = 0 \,.
\ee
The definition of the ghost number and the values of ghost numbers for
fields and antifields can be found in the Appendix (see relations
\rf{03112015-man03-37}, \rf{03112015-man03-38}). Using relations
\rf{16112015-man-01}, we find that BRST--BV Lagrangian
\rf{30102015-man03-04} takes the form
\be  \label{16112015-man-02}
\LL = \half \langle \phi_\Ism
| (\Box - M^{\oplussm\oplussm})
|\phi_\Ism \rangle
- \half \langle
\phi_\IIsm |
(\Box - M^{\oplussm\oplussm})
|\phi_\IIsm \rangle
- \langle \cb_*| \Lb_\Ism |\phi_\Ism\rangle
- \langle \cb_*| L_\IIsm |\phi_\IIsm\rangle  - \half \langle \cb_* |
|\cb_*\rangle\,,
\ee
where the operators $\Lb_\Ism$, $L_\IIsm$ appearing in
\rf{16112015-man-02} are defined in
\rf{28102015-man03-11-b1}--\rf{28102015-man03-11-b4}.  The equation of
motion for the antifield $|\cb_*\rangle$ obtained from Lagrangian
\rf{16112015-man-02} allows us to express the antifield
$|\cb_*\rangle$ in terms of the fields $|\phi_\Ism\rangle$,
$|\phi_\IIsm\rangle$ as
\be \label{18112015-man-03}
- |\cb_*\rangle =   \Lb_\Ism |\phi_\Ism\rangle  + | L_\IIsm |\phi_\IIsm\rangle\,.
\ee
Plugging \rf{18112015-man-03} in \rf{16112015-man-02}, we find that
Lagrangian \rf{16112015-man-02} takes the form of the Lagrangian in
the metric-like approach given in \rf{28102015-man03-11},
\rf{28102015-man03-11-b0}.

\noindent
\iiibf) Using the Siegel gauge $|\phi_*\rangle = 0$, we verify that
Lagrangian \rf{30102015-man03-04} leads to the following simple
expression for gauge-fixed Lagrangian:
\be \label{16082015-man-19-a1}
\LL
=
\half \langle \phi_\Ism
| (\Box - M^{\oplussm\oplussm})
|\phi_\Ism \rangle
- \half \langle
\phi_\IIsm |
(\Box - M^{\oplussm\oplussm})
|\phi_\IIsm \rangle
+
\langle
\cb |(\Box
- M^{\oplussm\oplussm})
|c\rangle\,.
\ee
We note that when passing from the relation for a gauge invariant
Lagrangian in \rf{30102015-man03-04} to the gauge-fixed Lagrangian in
\rf{16082015-man-19-a1}, we changed the sign of the Faddeev--Popov
antighost field, $\cbk\rightarrow -\cbk$. Gauge-fixed Lagrangian
\rf{16082015-man-19-a1} is invariant under the global BRST and
anti-BRST symmetries given by
\beq
\label{18112015-man-01} && \hspace{-1cm} \ssf  |\phi_{\Ism,\IIsm}\rangle   =     G_{\Ism,\IIsm} \ck\,, \qquad \ssf  \ck = 0\,, \hspace{4cm}  \ssf  \cbk   =  \Lb_\Ism |\phi_\Ism\rangle   + L_\IIsm |\phi_\IIsm\rangle  \,,
\\
\label{18112015-man-02} && \hspace{-1cm} \ssfb  |\phi_{\Ism,\IIsm}\rangle  =    G_{\Ism,\IIsm}  \cbk\,,\qquad\ssfb   \ck = -   \Lb_\Ism |\phi_\Ism\rangle - L_\IIsm |\phi_\IIsm\rangle\,,   \qquad \ssfb   \cbk   = 0 \,,
\eeq
where the operators $\Lb_\Ism$, $L_\IIsm$, and $G_{\Ism,\IIsm}$, are
respectively given in \rf{28102015-man03-11-b1}, \rf{28102015-man03-11-b2},
and \rf{30102015-man03-02-b1}, \rf{30102015-man03-02-b2}. We also note that the global BRST and
anti-BRST transformations presented in \rf{18112015-man-01},
\rf{18112015-man-02} satisfy the nilpotence relations $\ssf^2=0$,
$\ssfb^2=0$ for arbitrary Faddeev--Popov fields, while the nilpotence
relation $\ssf\ssfb + \ssfb\ssf=0$ is respected only for on-shell
Faddeev-Popov fields $\cbk$, $\ck$.

\newsection{  Realization of conformal symmetries  }\label{section-04}

Conformal symmetries of fields propagating in space-time $R^{d-1,1}$ are described by the conformal algebra $so(d,2)$. Note however that only symmetries of the Lorentz algebra $so(d-1,1)$ are manifest in the framework of our approach. Therefore in order to complete our description of conformal fields we should provide a realization of symmetries of the conformal algebra on space of fields and antifields entering the BRST--BV action. Because symmetries of the Lorentz algebra $so(d-1,1)$ are realized manifestly in the framework of our approach, we represent the $so(d,2)$ algebra in the basis of the Lorentz algebra  $so(d-1,1)$. In basis of the Lorentz algebra, the generators of the conformal algebra $so(d,2)$ are described by the Poincar\'e translation generators $P^a$, dilatation generator $D$, conformal boosts generators $K^a$, and generators of the Lorentz algebra $so(d-1,1)$ denoted by $J^{ab}$. We assume the following normalization for commutation relations of the $so(d,2)$ algebra generators
\beq
&& {}[D,P^a]=-P^a\,,
\hspace{2.5cm}
[P^a,J^{bc}]
=\eta^{ab}P^c
-\eta^{ac}P^b
\,,
\nonumber\\
&& [D,K^a]=K^a\,,
\hspace{2.7cm}
[K^a,J^{bc}]
=\eta^{ab}K^c
- \eta^{ac}K^b\,,
\\
&& [P^a,K^b]
= \eta^{ab}D
- J^{ab}\,,
\hspace{1.2cm}
[J^{ab},J^{ce}]
= \eta^{bc} J^{ae}
+ 3\hbox{ terms}.
\nonumber
\eeq

A general representation for the generators of the $so(d,2)$ algebra on space of conformal fields
is given by the following relations
\beq
\label{02112015-man03-01} && P^a = \partial^a\,,
\\
\label{02112015-man03-02} && J^{ab} = x^a\partial^b - x^b \partial^a + M^{ab}\,,
\\
\label{02112015-man03-03} && D = x^a\partial^a + \Delta\,,
\\
\label{02112015-man03-04} && K^a =  - \half x^b x^b \partial^a + x^a D + M^{ab} x^b + R^a\,. \qquad
\eeq
In \rf{02112015-man03-03}, the $\Delta$ stands for operator of conformal dimension, while, in \rf{02112015-man03-02}, \rf{02112015-man03-04}, the $M^{ab}$ stands for spin operator of the Lorentz algebra $so(d-1,1)$. The spin operator satisfies the commutation relations
\be
[M^{ab}, M^{ce}] = \eta^{bc} M^{ae} + 3 \hbox{ terms}.
\ee
For totally symmetric fields considered in this paper, the spin operator $M^{ab}$ takes the following well known form
\be
M^{ab} = \alpha^a\alphab^b - \alpha^b \alphab^a\,.
\ee

Operator $R^a$ appearing in \rf{02112015-man03-04} does not depend on coordinates $x^a$ of space-time $R^{d-1,1}$ and, in general, depends on the derivatives $\partial^a$. We note also that, in the standard higher-derivative approach to conformal fields, the operator $R^a$ is often equal to zero, while in the ordinary-derivative approach this operator turns out to be nontrivial. It is the finding of the operator $R^a$ that provides the real difficulty in the problem of realization of the conformal boost symmetries in the framework of the ordinary-derivative approach.

In this section, we present our result for a realization of the operators $\Delta$ and $R^a$ on space of the fields and antifields entering our ordinary-derivative BRST--BV formulation of conformal fields.
For the reader convenience, we start with the description of the operators $\Delta$ and $R^a$ for the metric-like double-traceless fields \cite{Metsaev:2007rw} and the metric-like traceless fields we discussed in Sec.\ref{sec-01} in this paper.

\noindent {\bf Operators $\Delta$ and $R^a$ for metric-like double-traceless fields}. Expression for the conformal dimension operator can easily be read from relation in \rf{28102015-man03-25}. Namely, using \rf{28102015-man03-25} and \rf{28102015-man03-18}, we see that realization of the operator $\Delta$ on space of the double-traceless ket-vector $\phik$ \rf{28102015-man03-18} takes the form
\be
\label{03112015-man03-01} \Delta   =    M^{\ominussm\oplussm} + \frac{d-2}{2}\,, \qquad
M^{\ominussm\oplussm} \equiv N_{\alpha^\ominussm} - N_{\alpha^\oplussm}\,.
\ee
Realization of the operator $R^a$ on space of the double-traceless ket-vector \rf{28102015-man03-18} is given by \cite{Metsaev:2007rw}
\be \label{03112015-man03-03}
R^a = - \half M^{\ominussm\ominussm} \partial^a +  M^{\ominussm a}\,,
\ee
where we use the notation
\beq
\label{03112015-man03-04} M^{\ominussm\ominussm} & \equiv &  4 \alpha^\ominussm \alphab^\ominussm\,,
\\
\label{03112015-man03-05} M^{\ominussm a}  & \equiv &    r_{0,1} \bar\alpha^a +  \Awt^a \rb_{0,1} \,,
\\
\label{03112015-man03-06} && \Awt^a  \equiv \alpha^a - \alpha^2 \frac{1}{2N_\alpha+d-2}\alphab^a\,,
\\
\label{03112015-man03-07} && r_{0,1} \equiv 2 \zeta e_\zeta \alphab^\ominussm\,, \qquad \rb_{0,1} \equiv - 2 \alpha^\ominussm e_\zeta \bar\zeta\,, \qquad e_\zeta \equiv  \Bigl(\frac{2s+d-4-N_\zeta}{2s+d-4-2N_\zeta}\Bigr)^{1/2}.\qquad
\eeq
The operator $\Awt^a$ appears in \rf{03112015-man03-05} because it is this operator that respects double-tracelessness constraint \rf{28102015-man03-17}. Namely, if a ket-vector $\phik$ satisfies double-tracelessness constraint \rf{28102015-man03-17}, then the following relation holds true: $(\alphab^2)^2\Awt^a \phik=0$.

\noindent {\bf Operators $\Delta$ and $R^a$ for metric-like traceless fields}. Realization of the operators $\Delta$ and $R^a$ on space of the traceless ket-vectors $|\phi_\Ism\rangle$, $|\phi_\IIsm\rangle$ can easily be obtained by using the realization of $\Delta$ and $R^a$ on space of the double-traceless ket-vector above-discussed.
To this end we use the interrelations between the double-traceless ket-vector $\phik$ and the traceless ket-vectors $|\phi_\Ism\rangle$, $|\phi_\IIsm\rangle$ given in \rf{28102015-man03-06} and \rf{28102015-man03-09}, \rf{28102015-man03-10}. Doing so, we see immediately that a realization of the operator $\Delta$ on space of the ket-vectors $|\phi_\Ism\rangle$, $|\phi_\IIsm\rangle$ takes the same form as before
\be
\label{03112015-man03-08} \Delta   =    M^{\ominussm\oplussm} + \frac{d-2}{2}\,, \qquad
M^{\ominussm\oplussm} \equiv N_{\alpha^\ominussm} - N_{\alpha^\oplussm}\,.
\ee
For the operator $R^a$, we get the following representation
\be \label{03112015-man03-09}
R^a = - \half M^{\ominussm\ominussm} \partial^a +  M^{\ominussm a}\,,
\ee
where the realization of the operator $M^{\ominussm\ominussm}$ on space of the ket-vectors $|\phi_\Ism\rangle$, $\phi_\IIsm\rangle$ takes the form
\be \label{03112015-man03-10}
M^{\ominussm\ominussm} = 4 \alpha^\ominussm \alphab^\ominussm\,,
\ee
while the realization of the operator $M^{\ominussm a}$ on space of the ket-vectors $|\phi_\Ism\rangle$, $\phi_\IIsm\rangle$ is given by
\beq
\label{03112015-man03-11} M^{\ominussm a} |\phi_\Ism \rangle  &  = &  (r_{0,1}\alphab^a + A^a \rb_{0,1}) |\phi_\Ism\rangle + 2 r_{0,1} A^a \NN |\phi_\IIsm\rangle\,,
\\
\label{03112015-man03-12} M^{\ominussm a} |\phi_\IIsm \rangle   & = &  \bigl( r_{0,1} u_{11\zeta } \alphab^a + A^a u_{11\zeta} \rb_{0,1}\bigr) |\phi_\IIsm\rangle
-   2 \NN \rb_{0,1} \alphab^a |\phi_\Ism\rangle\,,
\eeq
and, in relations \rf{03112015-man03-11}, \rf{03112015-man03-12}, we use the following notation:
\beq
\label{03112015-man03-14} && A^a  \equiv \alpha^a - \alpha^2 \frac{1}{2N_\alpha+d}\alphab^a\,,
\\
\label{03112015-man03-15} && r_{0,1} \equiv 2 \zeta e_\zeta \alphab^\ominussm\,, \qquad \rb_{0,1} \equiv - 2 \alpha^\ominussm e_\zeta \bar\zeta\,,\qquad e_\zeta \equiv \Bigl(\frac{2s+d-4-N_\zeta}{2s+d-4-2N_\zeta}\Bigr)^{1/2}\,,\qquad
\\
\label{03112015-man03-16} && u_{11\zeta}^{\vphantom{5pt}} \equiv \Bigl(\frac{(2s + d - 4 -2N_\zeta)(2s + d - 8 -2 N_\zeta)}{(2s+d-6-2N_\zeta)^2}\Bigr)^{1/2}\,,
\\
\label{03112015-man03-17} && \NN\equiv \bigl((2N_\alpha+d)(2N_\alpha+d-2)\bigr)^{-1/2}\,.
\eeq
We note that the operator $A^a$ \rf{03112015-man03-14} appears in relations  \rf{03112015-man03-11}, \rf{03112015-man03-12} because it is this operator that respects the tracelessness constraints \rf{28102015-man03-08}. Namely, taking into account that the ket-vectors $|\phi_\Ism\rangle$, $|\phi_\IIsm\rangle$ satisfy tracelessness constraints \rf{28102015-man03-08}, we verify that the following relations hold true: $\alphab^2 A^a |\phi_{\Ism,\IIsm}\rangle =0$.

\noindent {\bf Operators $\Delta$ and $R^a$ for BRST--BV fields and antifields}. We now describe our main result in this section. In our BRST--BV approach, fields and antifields are described by the ket-vector \rf{27-10-2015-manus03-01} subject to algebraic constraints \rf{27-10-2015-manus03-02}-\rf{27-10-2015-manus03-04}. We find the following realization of the conformal dimension operator $\Delta$ on space of ket-vector $\Phik$ \rf{27-10-2015-manus03-01}:
\beq
\label{03112015-man03-18} \Delta & = &  2 \theta\partial_\theta + M^{\eta\rho} +  M^{\ominussm\oplussm} + \frac{d-2}{2}\,,
\\
\label{03112015-man03-19} && M^{\ominussm\oplussm} \equiv N_{\alpha^\ominussm} - N_{\alpha^\oplussm}\,,
\\
\label{03112015-man03-20} && M^{\eta\rho} \equiv N_\eta - N_\rho\,.
\eeq
Comparing operator $\Delta$ \rf{03112015-man03-18} with the one for double-traceless and traceless fields in \rf{03112015-man03-01}, \rf{03112015-man03-08}, we note that, in our BRST--BV approach,  the conformal dimension operator $\Delta$ \rf{03112015-man03-18} depends not only on the operator $M^{\ominussm\oplussm}$ but also on the Grassmann coordinate $\theta$, the Grassmann derivative $\partial_\theta$ and the Grassmann odd oscillators entering the operators $N_\eta$, $N_\rho$ in \rf{03112015-man03-20}. Note also that, the operator $\Delta$ in \rf{03112015-man03-18}, when restricted to the space of ket-vectors $|\phi_\Ism\rangle$, $|\phi_\IIsm\rangle$ \rf{27-10-2015-manus03-06} entering BRST--BV ket-vector \rf{27-10-2015-manus03-06}, turns out to be equal to the one in metric-like approach  \rf{03112015-man03-08}, as it should be.

For the operator $R^a$, we find the following realization on space of ket-vector $\Phik$ \rf{27-10-2015-manus03-01}:
\be \label{03112015-man03-21}
R^a = - \half M^{\ominussm\ominussm} \partial^a +  M^{\ominussm a} + 2\theta M^{\rho a}\,,
\ee
where operators $M^{\ominussm\ominussm}$, $M^{\ominussm a}$, $M^{\rho a}$ appearing in \rf{03112015-man03-21} are defined by the following relations:
\beq
 \label{03112015-man03-22} M^{\ominussm\ominussm} & = & 4 \alpha^\ominussm \alphab^\ominussm\,,
\\
\label{03112015-man03-23} M^{\rho a} & = &
\rho
f_{\eta \zeta}
\alphab^a
+ A^a
\fb_{\rho \zeta}\rhob\,,
\\
\label{03112015-man03-24} && f_{\rho \zeta}
\equiv
\Bigl(\frac{2s
+d-4-2N_\zeta-2N_\rho}{2s
+ d-4-2N_\zeta}\Bigr)^{1/2} \,,
\\
\label{03112015-man03-25} && \fb_{\eta \zeta}
\equiv
\Bigl(\frac{2s
+ d-4-2N_\zeta-2N_\eta}{2s
+ d -4-2N_\zeta}\Bigr)^{1/2}\,,
\\
\label{03112015-man03-26} M^{\ominussm a} & = &
A^\zeta  \alphab^a
+ A^a \Ab^\zeta\,,
\\
&& A^a  \equiv \alpha^a - \alpha^2 \frac{1}{2N_\alpha+d}\alphab^a\,,
\\
\label{03112015-man03-27} && A^\zeta
\equiv \zeta \eb_{\eta\rho \zeta}^\ominussm
+ h_\zeta^\ominussm \rho \eta \zetab\,,
\\
\label{03112015-man03-28} && \Ab^\zeta \equiv   e_{\eta\rho \zeta}^\ominussm \zetab   +  \zeta \etab \rhob  \hb_\zeta^\ominussm\,,
\eeq
and, in relations \rf{03112015-man03-27}, \rf{03112015-man03-28}, we use the following notation
\beq
\label{03112015-man03-29} &&  \eb_{\eta\rho \zeta}^\ominussm  \equiv  2 e_\zeta u_{\eta\rho \zeta} \alphab^\ominussm\,,\qquad
e_{\eta\rho \zeta}^\ominussm  \equiv  - 2 e_\zeta u_{\eta\rho \zeta} \alpha^\ominussm\,, \qquad e_\zeta \equiv  \Bigl(\frac{2s+d-4-N_\zeta}{2s+d-4-2N_\zeta}\Bigr)^{1/2}\,,\qquad
\\
\label{03112015-man03-30} && u_{\eta\rho \zeta}
\equiv
\Bigl(\frac{(2s +d-4-2N_\zeta)(2s
+d-4-2N_\zeta-2N_\eta - 2N_\rho)}{(2s
+ d-4-2N_\zeta - 2N_\eta)(2s
+ d-4- 2N_\zeta - 2N_\rho)}\Bigr)^{1/2} \,,
\\
\label{03112015-man03-31} && h_\zeta^\ominussm  \equiv 4e_\zeta ((2s+d-4-2N_\zeta)(2s+d-6-2N_\zeta))^{-1/2} \alpha^\ominussm \,,
\\
\label{03112015-man03-32} && \hb_\zeta^\ominussm \equiv - 4 e_\zeta ((2s+d-4-2N_\zeta)(2s+d-6-2N_\zeta))^{-1/2} \alphab^\ominussm\,.
\eeq
From relations in \rf{03112015-man03-22}-\rf{03112015-man03-32}, we see that the operators $M^{\ominussm\ominussm}$, $M^{\ominussm a}$, $M^{\rho a}$ depend only on the oscillators. These relations provide the complete description of the operator $R^a$.

Using relations for the operators $M^{ab}$, $\Delta$, and $R^a$, we verify that BRST--BV action \rf{30102015-man03-03} is invariant under transformation of the conformal algebra symmetries $so(d,2)$ given in \rf{02112015-man03-01}-\rf{02112015-man03-04}. Using expressions for $\Delta$ and $R^a$ above described, we also verify the commutation relation for conformal boost generators, $[K^a,K^b]=0$.

\bigskip
\noindent {\bf Superalgebra $osp(d-1,1|2)$}. Some subset of the spin operators involved in our BRST--BV formulation of conformal field turns out to be related to the superalgebra $osp(d-1,1|2)$. In order to explain what just has been said let us consider the spin operators $M^{\eta a}$, $M^{\eta \eta}$, which enter the BRST operator \rf{30102015-man03-05}. Also let us consider the spin operators $M^{\eta\rho}$, $M^{\rho a}$, $M^{ab}$ which enter generators of the conformal symmetries. We now observe that, if to the spin operators $M^{\eta a}$, $M^{\eta \eta}$ entering the BRST operator and the spin operators  $M^{\eta\rho}$, $M^{\rho a}$, $M^{ab}$, entering the conformal symmetries, we add a new spin operator $M^{\rho\rho}$, which we define by the following expression $M^{\rho\rho} = \rho\rhob$, then we get the superalgebra $osp(d-1,1|2)$. Namely, we checked that the just mentioned spin operators indeed satisfy the commutation relations of the superalgebra $osp(d-1,1|2)$. For the reader convenience, we note that we use the following normalization for commutation relations of
the superalgebra $osp(d-1,1|2)$:
\beq
&& [M^{ab},M^{ce}] = \eta^{bc} M^{ae} + 3 \hbox{ terms},
\\[5pt]
&& [M^{\eta\rho},M^{\eta \eta}]
= 2
M^{\eta\eta}\,,
\nonumber\\
&& [M^{\eta\rho},M^{\rho \rho}]
= - 2
M^{\rho\rho}\,,
\nonumber\\
&& [M^{\eta\eta},M^{\rho\rho}]
= M^{\eta\rho}\,,
\\[5pt]
&& [M^{\eta\rho},M^{\eta a}]
= M^{\eta a}\,,
\nonumber\\
&& [M^{\eta\rho},M^{\rho a}]
=
- M^{\rho a}\,,
\nonumber\\
&& [M^{\eta\eta},M^{\rho a}]
=
M^{\eta a}\,,
\nonumber\\
&& [M^{\rho\rho},M^{\eta a}]
=
M^{\rho a}\,,
\\[5pt]
&& [M^{\eta a},M^{bc}]
= \eta^{ab}
M^{\eta c}
- \eta^{ac}
M^{\eta b} \,,
\nonumber\\
&& [M^{\rho a},M^{bc}]
= \eta^{ab}
M^{\rho c}
- \eta^{ac}
M^{\rho b}\,,
\\[5pt]
&& \{M^{\rho a}, M^{\eta b}\}
=
\eta^{ab}
M^{\eta\rho}
+ M^{ab}\,,
\nonumber\\
&& \{ M^{\eta a},M^{\eta b}\}
= - 2\eta^{ab}
M^{\eta\eta}\,,
\nonumber\\
&& \{ M^{\rho a},M^{\rho b}\}
=
2\eta^{ab}
M^{\rho\rho}\,.
\eeq
We recall also the superalgebra $osp(d-1,1|2)$ can be decomposed as
\beq
\underbrace{ M^{ab}}_{so(d-1,1)},\,
\underbrace{M^{\eta\rho},
M^{\eta\eta},
M^{\rho\rho}}_{sp(2)}\,\,,\,
\underbrace{M^{\eta a},
M^{\rho a}}_{\hbox{ coset }}\,.
\eeq

\newsection{ Conclusions } \label{section-05}

To summarize, in this paper, we developed a BRST--BV Lagrangian
formulation of arbitrary integer spin totally symmetric conformal
fields propagating in flat space. So far, the Lagrangian BRST--BV
formulation of arbitrary integer spin conformal fields was not
discussed in the literature.%
\footnote{ A higher-derivative Lagrangian for arbitrary integer spin
  conformal fields invariant under {\it global} BRST transformations
  was discussed in Ref.\cite{Metsaev:2014vda}. In the framework of the
  BRST--BV approach, equations of motion for arbitrary integer spin
  conformal fields were discussed in Ref.\cite{Bekaert:2012vt}. In the
  framework of the BRST--BV Lagrangian approach, study of a spin-2
  conformal field (Weyl gravity) can be found in
  Ref.\cite{Boulanger:2001he}.  }
We note, however, that in the earlier literature, the BRST approach
has been extensively used for studying nonconformal (massless and
massive) fields propagating in flat and AdS spaces (see, e.g.,
Refs.\cite{Bengtsson:1990un}--\cite{Metsaev:2015oza}). We think that
the use of methods developed in
Refs.\cite{Bengtsson:1990un}-\cite{Metsaev:2015oza} might be helpful
for better understanding various aspects of the BRST formulation of
conformal fields. In this paper, we studied totally symmetric
conformal fields. In
Refs.\cite{Vasiliev:2009ck,Marnelius:2008er,Metsaev:2008ba},
mixed-symmetry conformal fields were studied in the framework of a
metric-like approach. We expect that applications of the BRST approach
to the mixed-symmetry conformal fields might provide a simpler setup
for studying the mixed-symmetry conformal fields.  We note that in
this paper, we dealt with free conformal fields.  In the BRST approach
framework, many interesting methods have been developed for studying
interacting nonconformal (massless and massive) fields propagating in
flat and AdS spaces (see, e.g.,
Refs.\cite{Bekaert:2004dz}--\cite{Dempster:2012vw}).  The application
of those methods for studying conformal fields should lead to better
understanding the interacting conformal fields.  Needless to say, the
use of the BRST--BV approach for studying long conformal fields
propagating in flat space \cite{Metsaev:2014sfa,Metsaev:2015rda} and
short conformal fields propagating in the AdS space
\cite{Metsaev:2014iwa} seems to be fruitful directions to pursue.
Application of BRST approach to the study of algebraic aspects of
higher-spin algebras along the lines of \cite{Joung:2014qya} could
also be of some interest.

\bigskip {\bf Acknowledgments}. We thank A. Semikhatov for useful comments.
This work was supported by the RFBR Grant No.14-02-01172.

\setcounter{section}{0}\setcounter{subsection}{0}
\appendix{ \large Notation and conventions  }

We use mostly positive flat metric $\eta^{ab}=(-,+,\ldots,+)$. In scalar products, we drop the flat metric, $X^aY^a \equiv \eta_{ab}X^a Y^b$. Vector indices of the Lorentz algebra  $so(d-1,1)$ run over $a,b,c,e=0,1,\ldots ,d-1$.

Coordinates in space-time $R^{d-1,1}$ and Grassmann coordinate are denoted by $x^a$ and $\theta$ respectively. Derivatives of the space-time coordinates $x^a$ and left derivative of the Grassmann coordinate $\theta$ are denoted by $\partial^a \equiv \eta^{ab}\partial/\partial x^b$ and $\partial_\theta$ respectively. The normalization for the integral over $\theta$ is as follows $\int d\theta \theta =1$. Hermitian conjugation rules for the coordinates and the derivatives are defined as %
\be
(x^a,\theta)^\dagger = (x^a,\theta)\,, \qquad (\partial^a, \partial_\theta)^\dagger  = (-\partial^a,\partial_\theta).
\ee
For product of two operators $A$, $B$ having arbitrary ghost numbers, hermitian conjugation is defined according to the rule $(AB)^\dagger = B^\dagger A^\dagger$.

(Anti)commutation relations for creation operators $\alpha^a$, $\alpha^\oplussm$, $\alpha^\ominussm$, $\zeta$, $\eta$, $\rho$ and the respective annihilation  operators $\alphab^a$, $\alphab^\ominussm$, $\alphab^\oplussm$, $\zetab$, $\rhob$, $\etab$, the vacuum $|0\rangle$ and the hermitian conjugation rules are defined as
\beq
&& {} [\alphab^a,\alpha^b] = \eta^{ab},  \ \quad \ [\alphab^\oplussm,\alpha^\ominussm]=1, \ \qquad \ [\alphab^\ominussm,\alpha^\oplussm]=1,  \ \quad \ [\zetab,\zeta]=1, \ \quad
\nonumber\\
&& \{\rhob,\eta\}=1\,, \ \qquad \  \{\etab,\rho\} =1\,,\qquad
\\
&& \alphab^a |0\rangle = 0\,, \hspace{1.3cm} \alphab^\oplussm |0\rangle = 0\,,  \hspace{1.3cm} \alphab^\ominussm |0\rangle = 0\,,  \hspace{0.9cm} \zetab |0\rangle
 = 0\,,
\nonumber\\
&& \etab |0\rangle = 0\,, \hspace{1.5cm} \rhob |0\rangle = 0\,,
\\
\label{03112015-man03-39} && \alpha^{a \dagger} = \alphab^a\,, \hspace{1.4cm} \alpha^{\oplussm\dagger} = \alphab^\oplussm\,, \hspace{1.4cm} \alpha^{\ominussm\dagger} = \alphab^\ominussm\,, \hspace{1.1cm} \zeta^\dagger = \zetab\,,
\nonumber\\
&& \eta^\dagger = \etab\,, \hspace{1.8cm} \rho^\dagger = \rhob\,.
\eeq

The creation and annihilation operators are referred to as oscillators in this paper.
The oscillators $\alpha^a$, $\alphab^a$  transform as vectors of the Lorentz algebra $so(d-1,1)$, while the oscillators $\alpha^\oplussm$, $\alphab^\ominussm$, $\alpha^\ominussm$, $\alphab^\oplussm$, $\zeta$, $\zetab$, $\eta$, $\rhob$, $\rho$, $\etab$ transform as scalars of the Lorentz algebra $so(d-1,1)$.

Throughout this paper the following notation for the products of the derivatives and the oscillators is used
\beq
\label{03112015-man03-39-a1} && \hspace{-1.5cm} \Box \equiv \partial^a \partial^a \,,
\\
\label{03112015-man03-39-a2} &&  \hspace{-1.5cm} \alpar \equiv \alpha^a \partial^a\,, \hspace{1.2cm}\albpar \equiv \alphab^a \partial^a\,, \hspace{1.6cm} \alpha^2 \equiv \alpha^a \alpha^a\,, \hspace{1.4cm} \bar\alpha^2 \equiv \bar\alpha^a \bar\alpha^a\,,
\\
\label{03112015-man03-39-a3} && \hspace{-1.5cm} N_\alpha \equiv \alpha^a \alphab^a\,, \qquad \ \ N_{\alpha^\oplussm}\equiv \alpha^\oplussm \alphab^\ominussm  \,, \qquad \ \ \ \ N_{\alpha^\ominussm}\equiv \alpha^\ominussm \alphab^\oplussm  \,, \qquad \ \  N_\zeta \equiv \zeta \zetab \,,
\\
\label{03112015-man03-39-a4} && \hspace{-1.5cm} N_\eta \equiv \eta \rhob\,, \hspace{1.6cm}  N_\rho \equiv \rho \etab\,.\qquad
\eeq
Using \rf{03112015-man03-39}, we note the following hermitian conjugation rules for operators in \rf{03112015-man03-39-a3}, \rf{03112015-man03-39-a4}
\beq
&& \hspace{-1.5cm} N_\alpha^\dagger = N_\alpha\,, \qquad \ \ N_{\alpha^\oplussm}^\dagger =  N_{\alpha^\ominussm}  \,, \qquad \ \ N_{\alpha^\ominussm}^\dagger = N_{\alpha^\oplussm}\,, \qquad \ \ \ \  N_\zeta^\dagger = N_\zeta \,,
\\
&& \hspace{-1.5cm} N_\eta^\dagger = N_\rho\,, \hspace{1.1cm}  N_\rho^\dagger  = N_\eta\,.\qquad
\eeq

On space of ket-vector $\Phik$ \rf{27-10-2015-manus03-01}, the internal Faddeev-Popov ghost operator $N_\FPsm^\intrm$  is realized as
\be \label{03112015-man03-33}
N_\FPsm^\intrm = \theta\partial_\theta + N_\eta - N_\rho\,.
\ee
Using the notation $X$ for oscillator, we find the ghost number $q$ of the oscillator $X$ according to the relation $[N_\FPsm^\intrm,X]  = q X$. Using this relation, we get the following ghost numbers of $\theta$, $\partial_\theta$, and the oscillators:
\beq
\label{03112015-man03-34} && \gh(\alpha^a,\alpha^\oplussm,\alpha^\ominussm,\zeta) =0\,, \qquad \gh(\alphab^a,\alphab^\oplussm,\alphab^\ominussm,\zetab) =0\,,
\\
 \label{03112015-man03-35} && \gh(\theta, \eta, \rhob) = 1\,, \hspace{2cm} \gh(\partial_\theta, \etab, \rho) = -1\,.
\eeq

For ket-vectors given in \rf{27-10-2015-manus03-06},\rf{27-10-2015-manus03-07},  the ghost numbers  are defined as eigenvalues of the external Faddeev-Popov operator. Using the notation    $\Nbf_\FPsm^\ext$ for the external Faddeev-Popov operator, we note that eigenvalues of  $\Nbf_\FPsm^\ext$ are obtained by using the following relation:
\be \label{03112015-man03-36}
(N_\FPsm^\intrm + \Nbf_\FPsm^\ext)\Phik=0\,,
\ee
where the internal Faddeev-Popov operator $N_\FPsm^\intrm$ is defined in \rf{03112015-man03-33}.
Using \rf{03112015-man03-36}, the ghost numbers of ket-vectors given in \rf{27-10-2015-manus03-06}, \rf{27-10-2015-manus03-07} are found to be
\beq
\label{03112015-man03-37} && \gh(|\phi_\Ism\rangle)=0\,, \hspace{1cm} \gh(|c\rangle)=1\,, \hspace{1cm} \gh(|\cb\rangle)=-1\,, \hspace{1cm} \gh(|\phi_\IIsm\rangle)=0\,,
\\
\label{03112015-man03-38}  && \gh(|\phi_{\Ism *}\rangle)=-1\,, \hspace{0.5cm} \gh(|c_*\rangle)=-2\,, \hspace{0.5cm} \gh(|\cb_*\rangle)=0\,,  \hspace{1.2cm} \gh(|\phi_{\IIsm *}\rangle) = -1\,.\qquad
\eeq
The ghost numbers of the vacuum $|0\rangle$ is equal to zero. Taking this into account and using \rf{03112015-man03-34}, we conclude that ghost numbers \rf{03112015-man03-37} of ket-vectors \rf{27-10-2015-manus03-06} coincide with ghost numbers of tensor fields appearing on the right-hand side in relations \rf{27-10-2015-manus03-08}-\rf{27-10-2015-manus03-11}.
The same coincidence holds true for ghost numbers \rf{03112015-man03-38} of ket-vectors \rf{27-10-2015-manus03-07} and the corresponding ghost numbers  of scalar, vector, and tensor antifields which are obtainable by expanding antifields ket-vectors into the oscillators $\alpha^a$, $\alpha^\oplussm$, $\alpha^\ominussm$, $\zeta$.

Ghost numbers of the gauge transformation parameters can be obtained by using the following relation $(N_\FPsm^\intrm + \Nbf_\FPsm^\ext+1)\Xik=0$.

Throughout this paper we use the following hermitian conjugation rules for bra-vectors and ket-vectors: %
\beq
&& \Phibr= \Phik^\dagger,
\\
&&  \langle\phi_{\Ism,\IIsm}| = |\phi_{\Ism,\IIsm}\rangle^\dagger\,, \qquad  \quad  \ \ \cbr = \ck^\dagger, \qquad \quad  \quad \cbbr = - \cbk^\dagger\,,
\\
&& \langle\phi_{*\Ism,\IIsm}| = - |\phi_{*\Ism,\IIsm}\rangle^\dagger\,, \qquad  \langle c_*| = - |c_*\rangle^\dagger, \qquad \ \ \langle \cb_*| = |\cb_*\rangle^\dagger\,.
\eeq

\noindent {\bf Matching of terminology for ket-vectors in this paper and the standard terminology}. For the reader convenience, we now match terminology for ket-vectors \rf{27-10-2015-manus03-06}, \rf{27-10-2015-manus03-07} we use in this paper and the standard terminology used for those ket-vectors in the literature.  We recall that, in this paper, all ket-vectors in \rf{27-10-2015-manus03-06} are referred to as fields, while all ket-vectors in \rf{27-10-2015-manus03-07} are referred to as antifields. We note that such terminology does not coincide with the standard terminology in the literature.
According to the standard terminology, ket-vectors in \rf{27-10-2015-manus03-06}, \rf{27-10-2015-manus03-07}  having zero, positive, and negative ghost numbers are referred to as fields, ghost, and antifields respectively. Taking into account relations \rf{03112015-man03-37}, \rf{03112015-man03-38}, this implies that, according to the standard terminology, the ket-vectors $|\phi_\Ism\rangle$, $|\phi_\IIsm\rangle$, $|\cb_*\rangle$ in \rf{27-10-2015-manus03-06}, \rf{27-10-2015-manus03-07} are referred to as fields, the ket-vector $|c\rangle$ in \rf{27-10-2015-manus03-06} is referred to as ghost, while the ket-vectors $|\cb\rangle$, $|\phi_{*\Ism}\rangle$, $|c_*\rangle$, $|\phi_{*\IIsm}\rangle$ in \rf{27-10-2015-manus03-06}, \rf{27-10-2015-manus03-07} are referred to as antifields.

\end{document}